\documentclass[hyper]{JHEP}
\usepackage[latin1]{inputenc}
\usepackage{amsmath}
\usepackage{amsfonts}
\usepackage{amssymb}
\usepackage{graphicx}

\title{N spike D-strings in AdS Space with mixed flux}
\author{Sagar Biswas$^1$, Priyadarshini Pandit$^2$, Kamal L. Panigrahi$^2$\\
$^1$Department of Physics, Ramakrishna Mission Vidyamandira, Belur Math, Howrah 711 202, India\\
$^2$ Department of Physics, Indian Institute of Technology Kharagpur, Kharagpur 721 302, India\\

Email: \email{biswas.sagar@vidyamandira.ac.in,pandit006@iitkgp.ac.in, panigrahi@phy.iitkgp.ac.in}}
\abstract{We use Dirac-Born-Infeld action to study the spinning D-string in  $AdS_3 $ background in the presence of both NS-NS and RR fluxes. We compute the scaling relation between the energy (E) and spin (S) in the `long string limit'. The energy of these spiky string is found to be a function of spin with the leading logarithmic behaviour and the scaling relation appears to be independent of the amount of flux present.
We further discuss folded D-string solutions in $AdS_3$ background with pure NS-NS and R-R fluxes.}
		
\begin{document}
	\section{Introduction}
The AdS/CFT correspondence \cite{Maldacena:1997re}, \cite{Witten:1998qj}, \cite{Gubser:1998bc}, which is one of the exciting discoveries of the last two decades of modern theoretical physics, relates string theory or gravity theory on asymptotically Anti-de Sitter (AdS) space in $d$ dimensional bulk spacetime to a conformal field theory (CFT) living on its $(d-1)$ dimensional boundary. The most promising version of this duality states that $\mathcal{N}=4 $ Super Yang-Mills (SYM) theory with $SU(N)$ gauge group is equivalent to type IIB string theory on $AdS_5 \times S^5$. Verifying the conjectured duality for all generic values of parameters would require the knowledge of full spectrum of states in gravity side as well as all observables in the boundary field theory side, which is quite challenging. So one restricts to the domain of the correspondence to a particular limit, i.e. the planar limit where one can solve the strongly coupled dual field theory from the computable weak coupling perturbative behaviour of string theory. In the past few years, multiple attempts have been made to establish the duality in different sub-sectors of the theories on both sides of the duality. The semiclassical approximation is one of the most suitable methods which is used to find the string spectrum in a variety of target space geometries. The dispersion relation among various conserved charges of such strings (rotating, pulsating etc.) have been calculated in the `large charge limit' and were equated to the
anomalous dimensions of the dual boundary operators \cite{GKP}, \cite{Berenstein:2003gb}, \cite{spiky1}. In an attempt to understand the dual field theory operators, a large class of rigidly rotating and pulsating strings have been studied in various asymptotically AdS and non-AdS backgrounds \cite{Kluson:2008gf}, \cite{Biswas:2011wu}, \cite{Pradhan:2013sja}, \cite{Biswas:2014tia},  \cite{Panigrahi:2011be}, \cite{Panigrahi:2014sia}, \cite{Banerjee:2015nha}, \cite{Biswas:2013ela}, \cite{Panigrahi:2012bm}.

Apart from the usual $AdS_5/CFT_4$ duality, the other celebrated example of the gauge gravity correspondence in its lower dimensional form is the $AdS_3/CFT_2$ duality. In this case, the duality relates type IIB string theory on $AdS_3\times S^3\times T^4$ to the ${\cal N} = (4,4)$ superconformal field theory in D=3. In an attempt to understand the duality better, several semiclassical strings have been studied in this background as well 	\cite{Maldacena:2000hw}, \cite{Lee:2008sk}, \cite{VIII},  \cite{IX}, \cite{Abbott:2012dd}, \cite{Beccaria:2012kb}, \cite{Beccaria:2012pm}, \cite{Abbott:2013mpa}, \cite{Rughoonauth:2012qd}, \cite{Cardona:2014gqa}, \cite{Sfondrini:2014via}. More recently, the $AdS_3 \times S^3$ background in the presence of both R-R and NS-NS type fluxes ($H_3 = dB_2$ and $F_3 = dC_2$) has been proved to be integrable \cite{Cagnazzo:2012se}, \cite{Wulff:2014kja}. The integrable structure of this theory, including the $S$-matrix, has been discussed in 	\cite{HT}, \cite{BianchiHoare}, \cite{Borsato:2014hja},  \cite{B.Hoare}, \cite{FinitegapBabichenko}, \cite{Lloyd}, and a large number of classical string solutions have been extensively studied in \cite{BHT}, \cite{Rotating2}, \cite{HN1}, \cite{HN2}, \cite{multispike}. 
The background solution has been shown to satisfy type IIB supergravity field equations, provided the parameters associated to field strength of NS-NS fluxes ($q$) and the one associated to the strength
of R-R fluxes (say  $\hat{q}$) are related by the constraint $q^2+ {\hat q}^2= 1$. 
This model then interpolates between the AdS$_3$ background with R-R flux which can be studied by the integrability approach and the background with pure NS-NS flux which can be studied using the usual WZW model. To build the bridge, numerous string solutions have been studied in detail in this so called mixed flux background \cite{Rotating1}, \cite{both}, \cite{C.Ahn}, \cite{Banerjee:2018goh}.

Study of D-branes in $SL(2,\mathbb{R})$ Wess-Zumino-Witten (WZW) model, was among the pioneer attempts in employing D$p$-branes to establish the AdS/CFT dictionary \cite{Balog:1988jb}. Later giant magnon \cite{Hofman:2006xt} and single spike \cite{Ishizeki:2007we} like solutions were obtained from the D1-string rotating in $\mathbb{R} \times S^2$ and $\mathbb{R} \times S^3$ using Dirac-Born-Infeld (DBI) action \cite{Kluson:2007fr}. We have now a better understanding of the WZW D-branes from both the gravity side as well as the dual CFT side     \cite{Bachas:2000fr}, \cite{Banerjee:2016avv}, \cite{Kluson:2005mr}, \cite{Klimcik:1996hp}, \cite{Alekseev:1998mc}, \cite{Stanciu:2000fz}, \cite{Bachas:2000ik}, \cite{Alekseev:2000fd}. The integrability of D1 string on the group manifold with both NS-NS and RR fluxes is shown with the condition that Ramond-Ramond zero forms and dilaton should be constants only at the bosonic level \cite{Kluson:2015lia} . However, studying the dynamics of D1 string in this background, even in the bosonic sector continues to be a fascinating problem.  In the present paper, we study probe D1 string in $AdS_3$ background with mixed flux. In \cite{Banerjee:2019puc}, we had considered $N$-spike string solution in mixed flux with Jevicki-Jin \cite{Jevicki:2008mm},\cite{Jevicki:2009uz} type embedding and calculated the conserved charges to find the energy spin scaling relation in the large charge limit. The scaling relation for $N$-spike strings calculated from the sigma model using conformal gauge was given by
\begin{eqnarray}
E-S = N\frac{\sqrt{\lambda}}{\pi}\sqrt{1-q^2}\left[\frac{1}{2} {\rm log} S+ \cdots\right] \label{1.1}
\end{eqnarray}  
In this paper, we would like to study folded spinning D-string solutions by using the DBI action. Interestingly, we observe that $E-S$ scaling relation does not depend on the mixed flux parameter $q$. Though the result is somewhat strange, however apriori it is not clear what is to be expected from the spiky D1-string. In fact, there is no obvious relation between the F-string and D-string classical solutions 
(non-perturbatively there should be S-symmetry relation, but this does not apply to classical solutions). Furthermore, we discuss spiky D-strings in the presence of pure NS-NS and RR fluxes separately.\\

The rest of the paper is organised as follows. In section 2, we study the folded string solutions from the D1 string equations of motion in pure $AdS_3$ background. We find the energy-spin dispersion relation to be precisely same as was obtained in case of the F-string in $AdS_3$. Section 3 is devoted to the study of DBI action and the $N$ spiky strings in $AdS_3$ background in the presence of mixed R-R and NS-NS fluxes. We analyze the profile of the spiky strings in the presence of both the fluxes. We further find that the energy-spin dispersion relation is independent of the flux parameter $q$. We also comment on the pure NS-NS $(q = 1)$ and pure R-R flux $(q = 0)$ backgrounds. Finally, in section 4, we present our conclusion and outlook.

	\section{Spiky D1-strings in AdS space}
	
	To study the D1-string in generic background, we start with the DBI action which is given by,
	
	\begin{equation}
 		S = -T_1\int d\xi^0d\xi^1 e^{-\Phi} \sqrt{-\det A_{\alpha\beta}} + \frac{T_1}{2}\int d\xi^0d\xi^1 \epsilon^{\alpha\beta}C_{\alpha\beta},
	\label{2.1}\end{equation}
	where $T_1 = \frac{\sqrt{\lambda}}{2\pi g_s}$ is the D1-string tension, with $\lambda$ and $g_s$ being the  t'Hooft and string coupling constants respectively, $\Phi$ is the dilaton, $\xi^0$ and $\xi^1$ are the world-sheet coordinates,  $\epsilon^{\alpha \beta}$ is the antisymmetric tensor with $\epsilon^{01}=1$. The $C_{\alpha\beta}=\partial_\alpha X^\mu \partial_\beta X^\nu C_{\mu\nu}$ is the pull back of the RR field, and $A_{\alpha\beta} = G_{\alpha\beta} + B_{\alpha\beta} + 2\pi\alpha^{\prime} F_{\alpha\beta}$, where $G_{\alpha\beta} = \partial_{\alpha}X^M\partial_{\beta}X^N g_{MN}$ is the induced metric on the world volume, and $B_{\alpha\beta} = \partial_{\alpha}X^M\partial_{\beta}X^N b_{MN}$ is the pullback of the NS-NS flux and finally, $F_{\alpha\beta} = \partial_{\alpha}A_{\beta} - \partial_{\beta}A_{\alpha}$ is the field strength for the world volume gauge field.
	We start by writing down the metric for the $AdS_3$ background
	\begin{equation}
	ds^2 =~~ -\cosh^2\rho dt^2 + ~~d\rho^2 + ~~\sinh^2\rho d\theta^2
	\label{A}\end{equation}
We choose the following ansatz,
	\begin{eqnarray}
	t = \xi^0 \ , ~~~ \rho = \rho(\xi^1) \ , ~~~ \theta = \omega\xi^0 + \xi^1 \ .
	\label{B}\end{eqnarray}
   As the dilaton for pure $AdS_3$ background is zero, so we can take $e^{\Phi} = 1 $.  In this background, the Lagrangian density (\ref{2.1}) is given by,
	\begin{eqnarray}
	\mathcal{L} &=& -T_1 \sqrt{-\det A} \nonumber \\ &=& -T_1 \Big[\rho^{\prime 2} (\dot{t}^2 \cosh^2\rho - \dot{\theta}^2 \sinh^2\rho) + \theta^ {\prime 2} \sinh^2\rho \cosh^2\rho \Big]^{\frac{1}{2}},
	\end{eqnarray}
where $\dot{}$ and $^{\prime}$ denote the derivative with respect to $\xi^0$ and $\xi^1$ respectively. Now solving for the $\theta$ equation we get,
	\begin{equation}
	\frac{-T_1\sinh^2\rho\cosh^2\rho}{\sqrt{-\det A}}  =  C \ ,
	\end{equation}
	which gives,
	\begin{equation}
	\partial_1\rho = \frac{\sinh 2\rho \sqrt{\sinh^22\rho - \sinh^22\rho_0}}{2\sinh 2\rho_0 \sqrt{\cosh^2\rho - \omega^2\sinh^2\rho}} \ ,
	\end{equation}\\
		\begin{figure}[t]
		{\includegraphics[scale=1.0]{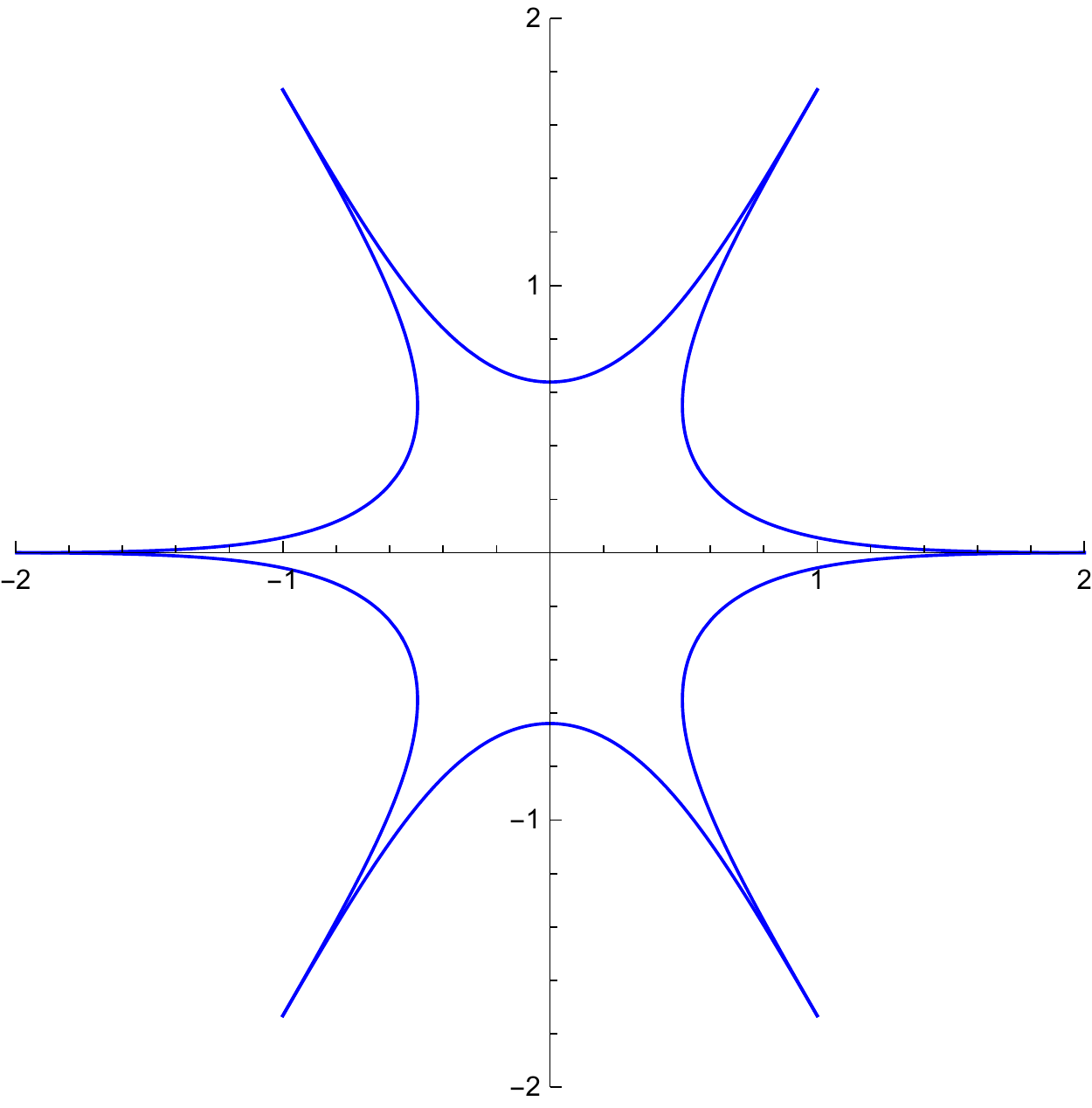}}
		\caption{ Spiky D-string which has 6 spikes in AdS space.} 
	\end{figure}
where $C$ is an integration constant, $\partial_1\rho= \frac{\partial \rho}{\partial \xi^1}$ and $\sinh 2\rho_0 = \frac{2C}{T_1 e^{-\Phi_0}}$. Now we can see from the expression for $\partial_1\rho$ that
	$\rho$ varies from a minimum value $\rho_0$ to a maximum value $\rho_1 = \coth^{-1}(\omega)$. At $\rho = \rho_1$, $\partial_1\rho$ diverges indicating the presence of a spike and at $\rho = \rho_0$, $\partial_1\rho$ vanishes, indicating the bottom of the valley between spikes. To get a solution with $N$ spikes we should glue $2N$ of the arc segments we get here. For that we need to choose $\rho_1$ and $\rho_0$ in such a way that the angle between cusp and valley is $\frac{2\pi}{2N}$. Fig. 1 shows the profile of six spike solutions of the D-string which precisely matches with the usual spiky strings in the context of AdS/CFT. To have better idea about these, it is imperative to have the scaling relationship among various charges. In what follows, we adopt the analysis of \cite{spiky1}. We make the following substitutions,
	\begin{eqnarray}
	u = \cosh 2\rho \ , ~~~ u_0 = \cosh 2\rho_0 \ , ~~~ u_1 = \cosh 2\rho_1 \ ,
	\end{eqnarray}
	and solving this we get,
	\begin{equation}
	\xi^1 = \frac{\sinh 2\rho_0}{\sqrt{2(u_1+u_0)}\sinh\rho_1} \Big[ \Pi(n_-, \alpha, k) - \Pi(n_+, \alpha, k) \Big],
	\label{C}\end{equation}
	where $\Pi$'s are the incomplete elliptic integrals of third kind with ~ $n_+=\frac{u_1 - u_0}{u_1 + 1}$, $n_-=\frac{u_1 - u_0}{u_1 - 1}$, $\sin\alpha = \sqrt{\frac{u_1 - u}{u_1 - u_0}}$ and $ k = \sqrt{\frac{u_1 - u_0}{u_1 + u_0}}$ .\\
	The conserved quantities like energy $E$ and spin $S$, can be computed as,
	\begin{eqnarray}
	E &=& 2NT_1\int_{\rho_0}^{\rho_1} \frac{\sinh\rho(4\cosh^4\rho - \omega^2\sinh^22\rho_0) d\rho}{2\cosh\rho \sqrt{\sinh^22\rho - \sinh^22\rho_0}\sqrt{\cosh^2\rho - \omega^2\sinh^2\rho}} \ , \nonumber \\ S &=& 2N\omega T_1\int_{\rho_0}^{\rho_1} \frac{\sinh\rho \sqrt{\sinh^22\rho - \sinh^22\rho_0} d\rho}{2\cosh\rho \sqrt{\cosh^2\rho - \omega^2\sinh^2\rho}} \ .  
	\end{eqnarray} 
	Note that $E$ and $S$ are multiplied by $2N$ to obtain the total
	energy and angular momentum. Now combining these we get the energy-spin relation $E-\omega S$ as,
\begin{equation}
    E - \omega S  =  2N T_1\int_{\rho_0}^{\rho_1} \frac{\sinh 2\rho \sqrt{\cosh^2\rho - \omega^2\sinh^2\rho}  d\rho}{\sqrt{\sinh^22\rho - \sinh^22\rho_0} } \ .
\end{equation}
 Again, using the above change of variables which is useful in determining the exact values of the conserved charges, we get\\
	\begin{eqnarray}
	E  & = & \frac{2NT_1\sinh\rho_1}{\sqrt{2(u_1 + u_0)}} \Big[(1 - u_0)K(k) + (u_1 + u_0)E(k) - \frac{\omega^2 (u_0^2 - 1)}{u_1+1}\Pi(n_+,k)\Big] \ , \nonumber \\ S  & = & \frac{2NT_1\cosh\rho_1}{\sqrt{2(u_1 + u_0)}} \Big[(u_1 + u_0)E(k)-(u_0 + 1)K(k) -  \frac{(u_0^2 -1)}{u_1+1}\Pi(n_+,k)\Big] \ , 
	\end{eqnarray}
and the energy-spin relation reduces to,
\begin{equation}
    E - \omega S = \frac{2NT_1\sqrt{u_1 + u_0}}{\sqrt{2}\sinh\rho_1}\Big[K(k) - E(k)\Big] \ , 
\end{equation}
	where $K(k)$, $E(k)$ and $\Pi(n_+,k)$s are the complete elliptic integrals of first, second and third kind respectively with the same value of $n_+$ and $k$ as stated earlier. Although it is interesting that the result can be expressed in terms of well-studied functions, the actual expressions are not very illuminating. To compare with the field theory we take the long string limit i.e. $\rho_1 >> \rho_0$. In this limit spin and energy-spin relation reduces to,
	\begin{eqnarray}
	S &=& 2NT_1 ~~\Big[\frac{\omega^2}{\omega^2 - 1}E(k) - K(k)\Big] , \nonumber \\ E - \omega S &=& 2NT_1 \omega ~~[K(k) - E(k)] \ .
	\end{eqnarray}
	Finally in infinite string limit $\omega \to 1$, using the expansion $E(k) = 1$ and $K(k) = \frac{3}{2} \log 2 - \frac{1}{2} \log(\omega - 1)$ we get,
	\begin{equation}
	S = 2NT_1 \Big[\frac{1}{\omega - 1} - \frac{3}{2}\log 2\Big] \ ,
	\end{equation}
	which shows $\log S = -\log(\omega - 1)$. Using this we get,
	\begin{equation}
	E - S = 2NT_1 \Big[\frac{1}{2} \log S + \cdots \Big] \ .
	\end{equation}
	If we substitute the D1-string tension $T_1 = \frac{\sqrt{\lambda}}{2\pi g_s}$, then the above dispersion relation looks like
	\begin{equation}
	E - S = \frac{N\sqrt{\lambda}}{\pi g_s} \Big[\frac{1}{2} \log S + \cdots \Big] \ . \label{1.2}
	\end{equation}
	Thus in infinite string limit, $S$ diverges and the leading order of $E-S$ have logarithmic divergences as found in \cite{Jevicki:2008mm}, \cite{Banerjee:2019puc}. The only difference is the constant scaling factor due to D1-string tension. This reproduces the well known dispersion relation of the folded spinning string or the GKP string for $N=2$.

\section{Spiky D1-string in AdS space with mixed flux}

	Let us now move on to discuss D1-string in the background of $AdS_3 \times S^3$ with mixed flux. The background, fluxes and the constant dilaton are given by

	\begin{equation*}
	ds^2 =~~ -\cosh^2\rho dt^2 + ~~d\rho^2 + ~~\sinh^2\rho d\theta^2
	\end{equation*}
	\begin{equation}
	 B_{t\theta} = q\sinh^2\rho , ~~~~~C_{t\theta}  =  \hat{q}\sinh^2\rho , ~~~~~\Phi  =  \Phi_0 ,
	\end{equation}
	where $\hat{q}=\sqrt{1-q^2}$. To see the individual effects of these fluxes in our solution, we use $q$ and $\hat{q}$ as the coefficient of NS-NS and RR fluxes respectively. The net effect of these fluxes can be seen by substituting $\hat{q}=\sqrt{1-q^2}$. 
	Using the same ansatz (\ref{B}), the Lagrangian density in the above background takes the following form,


\begin{eqnarray}
	\mathcal{L} &=& -T_1 e^{-\Phi_0}\Big[ \rho^{\prime 2} \big(\dot{t}^2\cosh^2\rho - \dot{\theta}^2 \sinh^2\rho \big) +\dot{t}^2 \theta^{\prime 2} \sinh^2 \rho \cosh^2 \rho - \dot{t}^2 \theta^{\prime 2} q^2\sinh^4\rho \Big]^{\frac{1}{2}} \nonumber \\ && +   T_1~\dot{t} \theta^{\prime} \hat{q} \sinh^2\rho \ .
\end{eqnarray}
Now solving equation of motion of $\theta$ we get,
	\begin{equation}
	\partial_1\rho=\frac{\sqrt{(\sinh^22\rho-4q^2\sinh^4\rho)\left[\sinh^22\rho-4q^2\sinh^4\rho-(\sinh2\rho_0-2e^{\Phi_0}\hat{q}\sinh^2\rho)^2 \right]}}{2\sqrt{\cosh^2\rho-\omega^2\sinh^2\rho}~~(\sinh2\rho_0-2e^{\Phi_0}\hat{q}\sinh^2\rho)} \ ,
	\end{equation}
	where again $\sinh 2\rho_0 = \frac{2C}{T_1 e^{-\Phi_0}}$. Also, we can see that at $\rho = \rho_1 = \coth^{-1}(\omega)$, $\partial_1\rho$ diverges indicating the presence of a spike. The `spike' positions as we can see remain unchanged but the position of the `valleys' are not the same due to addition of fluxes. The position of the valleys or minimum value of $\rho$ is found by setting $\partial_1\rho=0$, which gives two roots
	\begin{eqnarray}
   \sinh^2\rho_2 &= & \frac{-(1 + \hat{q}e^{\phi_0}\sinh 2\rho_0) + \sqrt{\cosh^22\rho_0-q^2\sinh^2\rho_0 + 2\hat{q} e^{\Phi_0}\sinh 2\rho_0}}{2(1-q^2 - \hat{q}^2e^{2\phi_0})} \nonumber \\  \sinh^2\rho_3 &= & \frac{-(1 + \hat{q}e^{\phi_0}\sinh 2\rho_0) - \sqrt{\cosh^22\rho_0-q^2\sinh^2\rho_0 + 2\hat{q} e^{\Phi_0}\sinh 2\rho_0}}{2(1-q^2 - \hat{q}^2e^{2\phi_0})}
	\end{eqnarray}
Before we get into the details of the solutions and their validity, few comments are in order regarding the effect of the mixed fluxes on the geodesics of such spiky strings. We plot the spiky D-string profiles in presence of both the fluxes. Here one can observe the slimming of the string profiles due to the combined effect of NS-NS and R-R  flux as compared to the profile of the F-string spike in the presence of NS-NS flux \cite{Banerjee:2019puc}. From this result one can conclude that the presence of RR flux reduces the fattening effect of the string profile caused due to NS-NS flux.

Note that $\rho_1$ is independent of both NS-NS and R-R fluxes, so this singularity will always be present whether the NS-NS and R-R fluxes are there or not. But, the zeros at $\rho_2$ and $\rho_3$ depends on both NS-NS and R-R fluxes and one of them should reduce to $\rho_0$ when both the NS-NS and RR fluxes were absent as in the previous section. One can identify that $\rho_2$ reduces to $\rho_0$ when one set $q=0$ and $\hat{q}=0$. So, the upper limit of integration remains the same as in the previous section but the lower limit of integration has been shifted from $\rho_0$ to $\rho_2$ due to presence of these fluxes. 
Looking for the string profile $\xi^1(u)$, we get,
\begin{figure}[t]
	{\includegraphics[scale=1.0]{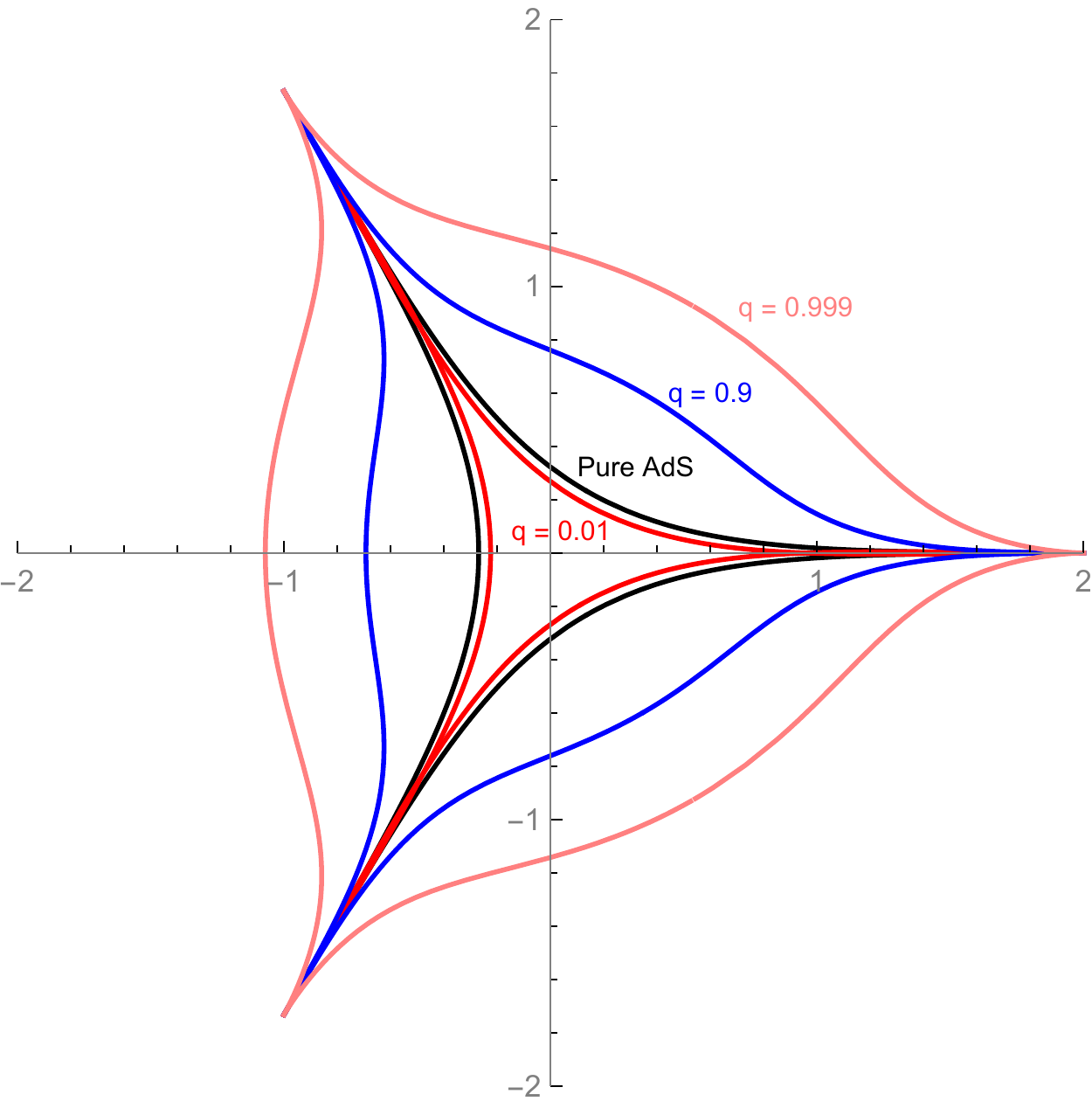}}
	\caption{Spiky D-string which has 3 spikes in AdS space in the presence of mixed fluxes. One can clearly observe the slimming effect of RR flux, i.e. in the presence of RR flux the fattening occured due to NS-NS flux get reduced. The spike profile for $q=0.01$ representing mostly due to RR flux is slimmer compared to spike profile in absence of any flux shown by pure AdS profile. Similarly, as depicted above, the slimming effect can be seen for other cases as well.}
\end{figure}
\begin{eqnarray}
    \xi^1 &=& \frac{1}{\sqrt{2(1-q^2-\hat{q}^2e^{2\Phi_0})}\sinh\rho_1} \times \nonumber \\ && \Big[\sinh 2\rho_0\int \frac{\sqrt{u_1-u} du}{(u-1)\sqrt{(1-q^2)u + (1+q^2)}\sqrt{(u+1)(u-u_2)(u-u_3)}} \nonumber \\ && - \hat{q} e^{\Phi_0} \int \frac{\sqrt{u_1-u} du}{\sqrt{(1-q^2)u + (1+q^2)} \sqrt{(u+1)(u-u_2)(u-u_3)}} \Big] \ ,
\end{eqnarray}
where 

\begin{eqnarray}
    u_2 &=& \frac{-[q^2 + \hat{q}^2e^{2\Phi_0} + \hat{q} e^{\Phi_0}\sinh 2\rho_0] + \sqrt{\cosh^2 2\rho_0 - q^2\sinh^2 2\rho_0 + 2\hat{q}e^{\Phi_0}\sinh 2\rho_0}}{(1-q^2 - \hat{q}^2e^{2\Phi_0})} \ , \nonumber \\ 
u_3 &=& \frac{-[q^2 + \hat{q}^2e^{2\Phi_0} + \hat{q}e^{\Phi_0}\sinh 2\rho_0] - \sqrt{\cosh^2 2\rho_0 - q^2\sinh^2 2\rho_0 + 2\hat{q}e^{\Phi_0}\sinh 2\rho_0}}{(1-q^2 - \hat{q}^2e^{2\Phi_0})} \ . \nonumber
\end{eqnarray} 
Now one can check that the integrations in the expressions of $\xi^1$ cannot be expressed in terms of any of the elliptic integrals any more. This complexity arises due to the presence of NS-NS flux. RR flux appears only in the coefficients and the zeros $u_2$ and $u_3$, hence it doesnot affect the complexity of the integrations. One can also check that if we put $q=0$ (pure RR flux) then these integrations can be expressed as elliptic integrals as can be seen in subsection (3.1). Also, using $\hat{q}=0$ or $q=1$ (pure NS-NS flux) not only makes the coefficient of the second integration zero but also reduces the complexity of the integration and makes it expressible in terms of elliptic integrals as can be seen in subsection (3.2). Thus only for $q=0$ and $q=1$ the integrations can be expressed in terms of elliptic integrals.\\
For any general value of $q$ (other than 0 and 1), to slove the integrations analytically we need a more general kind of integrals which will reduce to the elliptic integrals under $q=0$ and $q=1$. In terms of these generalised integrals we have,
\begin{eqnarray}
    \xi^1 &=& \frac{\sqrt{2}}{\sqrt{1-q^2-\hat{q}^2e^{2\Phi_0}} \sqrt{1+q^2+(1-q^2)u_1}\sqrt{u_1-u_3}\sqrt{u_1+1} \sinh\rho_1} \times \nonumber \\  && \Big[ \sinh 2\rho_0 I_2(\alpha,q,k)  + \hat{q}e^{\Phi_0}(u_1+1)F_1(\alpha,q,k) \nonumber \\ && - [\sinh 2\rho_0 + \hat{q}e^{\Phi_0}(u_1+1)] \Pi(n_+,\alpha,q,k) \Big] \ , 
\end{eqnarray}
where $F_1(\alpha,q,k)$, $\Pi(n_+,\alpha,q,k)$ and $I_2(\alpha,q,k)$ are the incomplete generalised integrals defined in Appendix A with $n_+ = \frac{u_1 - u_2}{u_1+1}$, $n_- = \frac{u_1 - u_2}{u_1-1}$, $\sin\alpha = \sqrt{\frac{u_1-u}{u_1-u_2}}$, $k^2 = \frac{u_1-u_2}{u_1-u_3}$ and $p^2 = \frac{(1-q^2)(u_1-u_2)}{1+q^2 + (1-q^2)u_1}$.\\

\noindent 
Now one can note that as $q \to 0$,~ $p^2 \to n_+$~ and under these conditions,\\ $F_1(\alpha,q,k) \to F(\alpha,k)$, $\Pi(n_+,\alpha,q,k)~ \to~ \Pi(n_+,\alpha,k)$ ~and\\ $I_2(\alpha,q,k) \to \frac{1}{2}[(u_1 + 1)\Pi(n_-,\alpha,k) - (u_1-1)\Pi(n_+,\alpha,k)]$. 

Thus, all the incomplete generalised integrals reduces to the incomplete elliptic integrals of different kinds.
Now let us try to calculate the spin $S$,
\begin{equation} 
S=NT_1e^{-\Phi_0}\omega\int \frac{\sinh\rho\sqrt{\sinh^22\rho - 4q^2\sinh^4\rho - \left(\sinh 2\rho_0 - 2\hat{q}e^{\Phi_0} \sinh^2\rho\right)^2} }{\sqrt{(\cosh^2\rho - \omega^2\sinh^2\rho)(\cosh^2\rho - q^2\sinh^2\rho)}}d\rho
\end{equation}
and the energy-spin relation $E-\omega S$,
\begin{eqnarray}
    && E-\omega S = 4NT_1e^{-\Phi_0} \int \frac{\sinh\rho \sqrt{(\cosh^2\rho - \omega^2\sinh^2\rho)(\cosh^2\rho - q^2\sinh^2\rho)} d\rho}{\sqrt{\sinh^22\rho - 4q^2\sinh^4\rho - (\sinh 2\rho_0 - 2\hat{q}e^{\Phi_0} \sinh^2\rho)^2}} \nonumber \\  && - 2NT_1\hat{q}  \int \frac{\sinh\rho[\sinh 2\rho_0 - 2\hat{q}e^{\Phi_0} \sinh^2\rho] \sqrt{\cosh^2\rho - \omega^2\sinh^2\rho} d\rho}{\sqrt{\cosh^2\rho - q^2\sinh^2\rho} \sqrt{\sinh^22\rho - 4q^2\sinh^4\rho - (\sinh 2\rho_0 - 2\hat{q}e^{\Phi_0} \sinh^2\rho)^2}} \nonumber \\
\end{eqnarray}
One can again note that the integrations of these expressions also can't be expressed in terms of the elliptic integrals and the NS-NS flux is responsible for this complexity. 
But, in terms of the generalised integrals we have,
\begin{eqnarray}
    S &=&  \frac{\sqrt{2}N\omega T_1 e^{-\Phi_0}\sqrt{1-q^2-\hat{q}^2e^{2\Phi_0}} \sqrt{u_1-u_3}\sinh\rho_1}{\sqrt{1+q^2+(1-q^2)u_1} \sqrt{u_1+1}} \times \nonumber \\ && \Big[ (u_1+1)E_1(q,k) - (1+u_2)I_1(q,k)\Big] \ , \label{3.3}
\end{eqnarray}
and
\begin{eqnarray}
     E - \omega S &=& \frac{\sqrt{2}NT_1e^{-\Phi_0}\sqrt{1+q^2+(1-q^2)u_1} \sqrt{u_1-u_3}}{\sqrt{1-q^2 -\hat{q}^2e^{2\Phi_0}}\sqrt{u_1+1}\sinh\rho_1} \Big[K_2(q,k) - E_2(q,k)\Big] \nonumber \\ && - \frac{\sqrt{2}NT_1\hat{q} \sqrt{u_1+1}}{\sqrt{1-q^2 -\hat{q}^2e^{2\Phi_0}} \sqrt{1+q^2+(1-q^2)u_1} \sqrt{u_1-u_3}\sinh\rho_1} \times \nonumber \\ && \Big[ (\sinh 2\rho_0 + 2\hat{q}e^{\Phi_0})\Pi(n_+,q,k) + \hat{q}e^{\Phi_0}(u_1-u_3)E_1(q,k) \nonumber \\ && - [\sinh 2\rho_0 + \hat{q}e^{\Phi_0}(2+u_1-u_3)K_1(q,k)]\Big] \ , \label{3.4}
\end{eqnarray}
where $K_1(q,k)$, $K_2(q,k)$, $E_1(q,k)$, $E_2(q,k)$, $\Pi(n_+,q,k)$, and $I_2(q,k)$ are the complete generalised integrals defined in Appendix A.
Now, one can once again note that as $q \to 0$, $p^2 \to n_+$ and under these considerations 
\begin{eqnarray}
K_1(q,k), K_2(q,k) \to K(k),\nonumber \\ E_1(q,k), E_2(q,k) \to E(k), \nonumber \\ \Pi(n_+,q,k) \to \Pi(n_+,k) \ , \nonumber \\
I_1(q,k) \to \frac{1}{u_1-u_3}\left[(u_1+1) K(k)-(u_3+1)\Pi(n_+,k)\right].
\nonumber 
\end{eqnarray}
Thus all the complete generalised integrals reduced to complete elliptic integrals of different kinds. \\

\noindent

\begin{figure}[t]
{\includegraphics[scale=0.4]{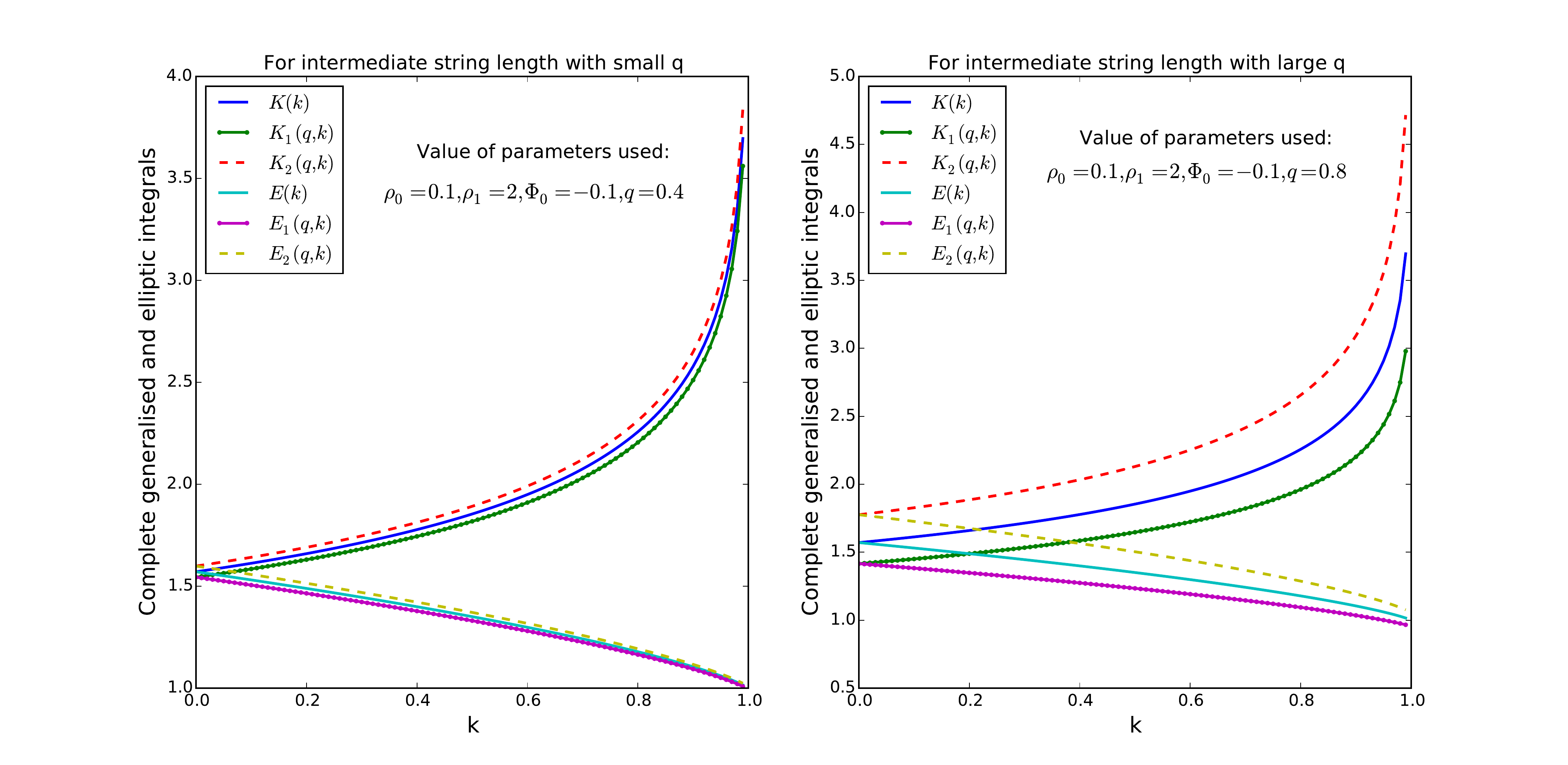}}
\caption{Complete generalised integrals $K_1(q,k)$, $K_2(q,k)$, $E_1(q,k)$ and $E_2(q,k)$ has been plotted alongwith complete elliptic integrals of first kind $K(k)$ and second kind $E(k)$ for intermediate length of string with $q=0.4$ and $q=0.8$. It is clearly showing that the curve for $K(k)$ is splitted into lower $K_1(q,k)$ and upper $K_2(q,k)$ curves, similarly the curve for $E(k)$ has been splitted into lower $E_1(q,k)$ and upper $E_2(q,k)$ curves. The splitting increases with NS-NS flux.}\label{fig1}
\end{figure}

\begin{figure}[t]
{\includegraphics[scale=0.4]{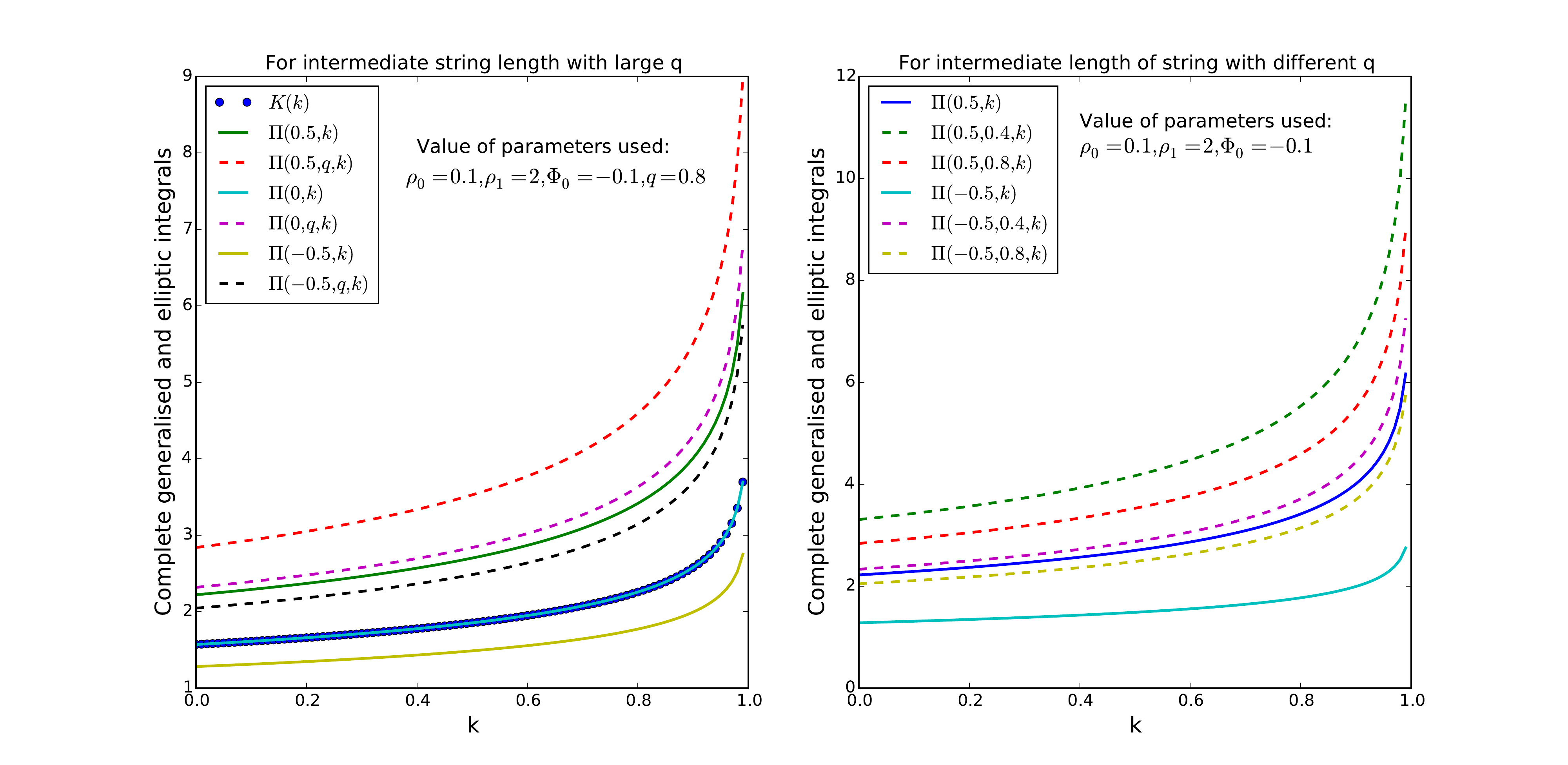}}
\caption{Complete generalised integral $\Pi(n_+,q,k)$ has been compared with complete elliptic integrals of third kind $\Pi(n_+,k)$ under intermediate string length for (i) fixed NS-NS flux with different $n_+$ and (ii) with different NS-NS flux. The first plot shows that $\Pi(n_+,q,k)$ has always higher value than $\Pi(n_+,k)$ for a particular $n_+$. The second plot shows that with increasing $q$ the curve $\Pi(n_+,q,k)$ comes closer to $\Pi(n_+,k)$.} \label{fig3}
\end{figure} 

\begin{figure}[t]
{\includegraphics[scale=0.4]{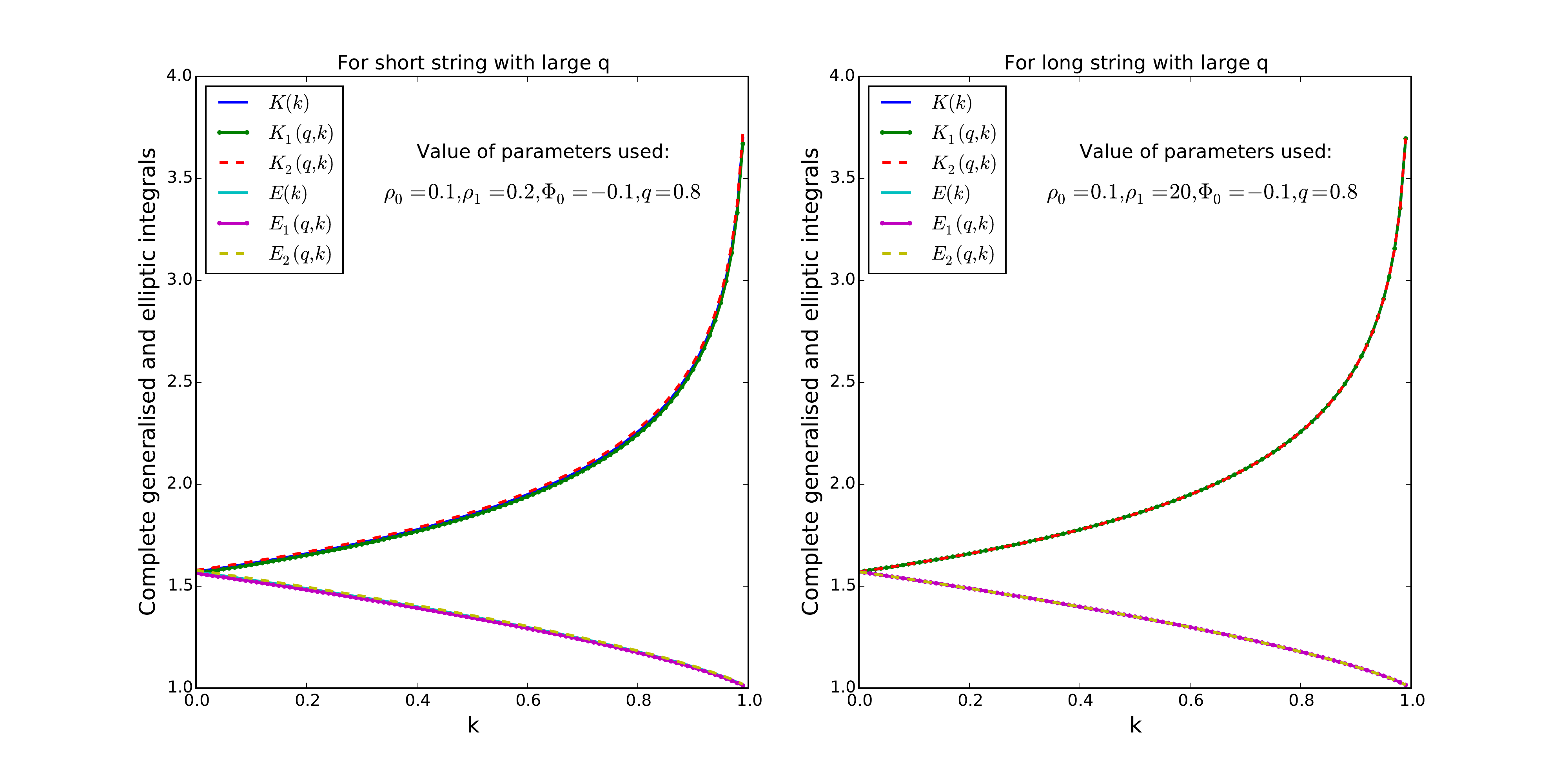}}
\caption{The same complete generalised integrals has been plotted alongwith complete elliptic integrals under short string and long string limit with $q=0.8$. Although, under short string limit both $K_1(q,k)$, $K_2(q,k)$ and $E_1(q,k)$, $E_2(q,k)$ are slightly off the track from the curves of $K(k)$ and $E(k)$ respectively, but for long string limit both of them converges to them irrespective of the value of NS-NS flux.}\label{fig2}
\end{figure}

As all the incomplete and complete generalised integrals reduce to incomplete and complete elliptic intetgrals of different kinds for $q=0$, we can think of these as the generalisation of the elliptic integrals. One can note that we have two types of generalisation of the elliptic integrals of first and second kind. 
In order to study how these generalised integrals behave we have integrate them numerically and plotted them as a function of $k$ as shown in Figure \ref{fig1} to Figure \ref{fig2}. Although there is no clear distinction but we have plotted them for mainly three different lengths of string and we discuss the salient features of them as follows:
\begin{enumerate}
    \item {\bf String of intermediate length:} For this case we consider the ratio $\rho_1/\rho_0 \sim 20$. 

(i) Complete elliptic integral of first kind $K(k)$ starts with a initial value (at $k=0$) and diverges at $k=1$. The generalised integrals $K_1(q,k)$ and $K_2(q,k)$ show similar type of behaviour. But, they start with a different initial value. Starting with a lower initial value $K_1(q,k)$ always keep its value lower than $K(k)$ until it diverges at $k=1$. Similarly $K_2(q,k)$ starts with a higher initial value and always keep its value higher than $K(k)$ until its diverging point. Thus it seems that the initial points determines the fate of the curve it follows. One can also note that with the increase of NS-NS flux the initial values spread ($K_1(q,0)$ have more lower value while $K_2(q,0)$ have more higher value) and as a result spreading of these curves increases as shown in Figure \ref{fig1}.

(ii) The complete elliptic integral of second $E(k)$ starts with the same initial value as $K(0)$ but they becomes to unity at $k=1$. The generalized integrals $E_1(q,k)$ and $E_2(q,k)$ also behave like $E(k)$ with different initial values. One can note that the initial value $E_1(q,0)$ is same as $K_1(q,0)$ and the initial value  $E_2(q,0)$ is same as $K_2(q,0)$. This suggest that if $E(k)$ is second kind of elliptic integrals whose first kind is $K(k)$ then $E_1(q,k)$ and $E_2(q,k)$ are the second kind of generalised integrals whose first kinds are $K_1(q,k)$ and $K_2(q,k)$ respectively. Also the spreading of $E_1(q,k)$ and $E_2(q,k)$ increases with NS-NS flux alongwith the initial points.

(iii) The complete elliptic integrals of third kind $\Pi(n_+,k)$ follow the same curve as $K(k)$ for $n_+=0$ but $\Pi(n_+,q,k)$ starts with a higher initial value and diverges at $k=1$ with its value always higher than $\Pi(0,k)$. For other non-zero values $n_+$, it is found that $\Pi(n_+,q,k)$ starts with a higher initial value than $\Pi(n_+,k)$ and follows a higher value curve as can be seen in first plot of Figure \ref{fig3}. From the second plot of Figure \ref{fig3} it is clear that as NS-NS flux increases the curve for $\Pi(n_+,q,k)$ comes closer to $\Pi(n_+,k)$ for a particular value of $n_+$. Thus unlike $K_{1,2}(q,k)$ and $E_{1,2}(q,k)$ here spreading decreases with increasing NS-NS flux.
    \item {\bf Short String:} For short string we consider the ratio $\rho_1/\rho_0 \sim 2$. As it can be seen from the first plots Figure \ref{fig2} that the curves of $K_1(q,k)$ and $K_2(q,k)$ are nearly touching the curve of complete elliptic integral of first kind $K(k)$. Also the curves of $E_1(q,k)$ and $E_2(q,k)$ are nearly touching the curve of $E(k)$. Thus the initial values and the spreading of the curves not only depends on the NS-NS fluxes but also on the length of the string we consider. 
\item {\bf Long String:} For long string we consider the ratio $\rho_1/\rho_0 \sim 200$ and under this consideration $K_1(q,k)$ and $K_2(q,k)$ is found to converge to $K(k)$ also $E_1(q,k)$ and $E_2(q,k)$ is found to converge to $E(k)$ as shown in the second plot of Figure \ref{fig2}. One can check the similar kind of behaviour for $\Pi(n_+,k)$ and $\Pi(n_+,q,k)$. Thus, we conclude that the effect of the fluxes is also not visible under long string limit. 
\end{enumerate}

Although under short string limit small effect of NS-NS flux can be seen, but under long string limit the effects of this flux is not visible at all. This indicates that under long string limit the above complicated expressions of spin and energy-spin relation might get simplified. Following the previous section we now use the long string limit $\rho_1>>\rho_0$, for this the arguments of the generalised integrals reduces as $n_+ \to \frac{1}{\omega^2}$, $k^2 \to \frac{1-q^2 - \hat{q}^2e^{2\Phi_0}}{\omega^2-q^2 - \hat{q}^2e^{2\Phi_0}}$ and $p^2 \to  \frac{1-q^2}{\omega^2-q^2}$. Under these conditions spin $S$ reduces to,
\begin{equation}
    S = \frac{2NT_1e^{-\Phi_0} \sqrt{\omega^2 - q^2 - \hat{q}^2e^{2\Phi_0}}}{\sqrt{\omega^2 - q^2}} \Big[ \frac{\omega^2}{\omega^2 - 1} E_1(q,k) - I_1(q,k)\Big] \ ,
\end{equation}
and energy-spin relation $E-\omega S$ becomes,
\begin{eqnarray}
    E &- &\omega S = \frac{2NT_1e^{-\Phi_0} \sqrt{\omega^2 - q^2} \sqrt{\omega^2 - q^2 - \hat{q}^2e^{2\Phi_0}}}{\omega (1 - q^2 - \hat{q}^2e^{2\Phi_0})} \Big[ K_2(q,k) - E_2(q,k)\Big] \nonumber \\ && - \frac{2NT_1\omega \hat{q}^2e^{\Phi_0} }{\sqrt{\omega^2 - q^2} \sqrt{\omega^2 - q^2 - \hat{q}^2e^{2\Phi_0}}} \Big[\frac{\omega^2 - q^2 - \hat{q}^2e^{2\Phi_0}}{1 - q^2 - \hat{q}^2e^{2\Phi_0}}E_1(q,k) \nonumber \\ && - \frac{[\omega^2 - 1 + \omega^2(1 - q^2 - \hat{q}^2e^{2\Phi_0})]}{1 - q^2 - \hat{q}^2e^{2\Phi_0}} K_1(q,k) + (\omega^2-1)\Pi(n_+,q,k) \Big] \ .
\end{eqnarray}

	Although the above expressions simplifies considerably but the complete generalised integrals still didn't get reduced to the complete elliptic integrals. To acheive that we take the infinite string limit i.e., $\omega \to 1$. One can note that in this limit, $n_+ = p^2 = 1$, and hence the above complete generalised integrals reduce to complete elliptic integrals. Thus in terms of these complete elliptic integrals we have,
\begin{equation}
    S = \frac{2NT_1e^{-\Phi_0} \sqrt{1 - q^2 - \hat{q}^2e^{2\Phi_0}}}{\sqrt{1 - q^2}} \Big[ \frac{1}{2(\omega - 1)} E(k) - K(k)\Big] \ ,
\end{equation}
and 
\begin{equation}
    E-S = \frac{2NT_1e^{-\Phi_0} [1 - q^2 + \hat{q}^2e^{2\Phi_0}]}{\sqrt{1 - q^2} \sqrt{1 - q^2 - \hat{q}^2e^{2\Phi_0}}} \Big[ K(k) - E(k)\Big] \ .
\end{equation}
Expanding $E(k)$ and $K(k)$ we find that $\log S = -\log(\omega - 1)$ and substituting the D1-string tension $T_1 = \frac{\sqrt{\lambda}}{2\pi g_s}$, the energy-spin dispersion relation becomes,
\begin{equation}
    E-S = \frac{N\sqrt{\lambda}e^{-\Phi_0} [1 - q^2 + \hat{q}^2e^{2\Phi_0}]}{\pi g_s\sqrt{1 - q^2} \sqrt{1 - q^2 - \hat{q}^2e^{2\Phi_0}}} \Big[ \frac{1}{2}\log S + \cdots \Big] \ . \label{3.5}
\end{equation} 

Thus, in the infinite string limit ($\omega \to 1$) we see that the complexity arose in the integrals due to the presence of the NS-NS fluxes have been removed. Also, we can see here that $S$ diverges and the leading order of $E-S$ have log divergence as in the case of no flux conditions. But, the individual contribution of these fluxes are still seen in the coefficient in the above relation.
 
	Finally, the net effect of these fluxes can be obtained by substituting $\hat{q} = \sqrt{1-q^2}$, the energy-spin relation turns out to be:

 \begin{equation}
    E-S = \frac{N\sqrt{\lambda}e^{-\Phi_0} [1  + e^{2\Phi_0}]}{\pi g_s \sqrt{1 - e^{2\Phi_0}}} \Big[ \frac{1}{2}\log S + \cdots \Big] \ .
\label{3.1}\end{equation}
	
 In the above relation (\ref{3.1}), it is very surprising to note that the dispersion relation is scaled by a factor which is independent of any parameters of flux present but the $log~S$ behaviour in the leading order remains intact as we obtain for the case of F-string in pure AdS$_3$. Infact when one analyzes a similar solution from the F-string in conformal gauge in AdS$_3$ with mixed flux, the coupling constant is scaled by $\sqrt{1-q^2}$ that appeared in the energy-spin dispersion relation \cite{Banerjee:2019puc}. Apriori one would expect a similar result in case of D-string solution as well because of S-duality. However, we note that there is no obvious relation between  the F-string and D1-string  at the level of classical  solutions, even though non-perturbatively there  should be S-duality symmetry relation between these two.

For $N=2$, equation (\ref{3.1}) becomes
	 \begin{equation}
	E-S = \frac{\sqrt{\lambda}e^{-\Phi_0} [1  + e^{2\Phi_0}]}{\pi g_s \sqrt{1 - e^{2\Phi_0}}} \Big[ \log S + \cdots \Big] \ .
\label{3.2}	\end{equation}
this gives the dispersion relation of the well known GKP string along with a constant factor which is independent of flux parameters. 

\subsection{Case of $q=0$: Pure R-R flux}
	We consider here the background supported purely by RR flux and no NS-NS flux. This can be obtained by putting $q=0$. So, the Lagrangian density can be written as,
\begin{eqnarray}
	\mathcal{L} =  -T_1 e^{-\Phi_0}\Big[ \rho^{\prime 2} \big(\dot{t}^2\cosh^2\rho &-& \dot{\theta}^2 \sinh^2\rho \big) +\dot{t}^2 \theta^{\prime 2} \sinh^2 \rho \cosh^2 \rho \Big]^{\frac{1}{2}} \nonumber \\ && + T_1~\dot{t} \theta^{\prime} \hat{q} \sinh^2\rho. 
\end{eqnarray}
Now solving the $\theta$ equation we get,
\begin{equation}
	\partial_1\rho = \frac{\sinh 2\rho \sqrt{\sinh^22\rho - (\sinh 2\rho_0 - 2e^{\Phi_0}\sinh^2\rho)^2}}{2 \sqrt{\cosh^2\rho - \omega^2\sinh^2\rho}[\sinh 2\rho_0 - 2e^{\Phi_0} \sinh^2\rho]} \ ,
\end{equation}
	where $\sinh 2\rho_0 = \frac{2C}{T_1 e^{-\Phi_0}}$. Now we can see from the expression for $\partial_1\rho$ that the spike is still at $\rho_1 = \coth^{-1}(\omega)$. To find the minimum value of $\rho$ we set, $\partial_1\rho = 0$, and changing the variables from $\rho$ to $u$ as before, we get two roots,
\begin{eqnarray}
	u_2 &=& \frac{-e^{\Phi_0}(e^{\Phi_0} + \sinh 2\rho_0) + \sqrt{\cosh^2 2 \rho_0 + 2e^{\Phi_0}\sinh 2\rho_0}}{1 - e^{2\Phi_0}} \ , \nonumber \\  u_3 &=& \frac{-e^{\Phi_0}(e^{\Phi_0} + \sinh 2\rho_0) - \sqrt{\cosh^2 2\rho_0 + 2e^{\Phi_0}\sinh 2\rho_0}}{1 - e^{2\Phi_0}} \ .
\end{eqnarray} 
One can once again identify that among $u_2$ and $u_3$, $u_2$ is the bottom of the valley between spikes. 
Under this condition the string profile reduces to,
\begin{eqnarray} 
	\xi^1 &=& \frac{1}{\sqrt{2(u_1-u_3)}\sqrt{1 - e^{2\Phi_0}}\sinh\rho_1} \Big[ 2e^{\Phi_0} F(\alpha,k) + \sinh 2\rho_0 \Pi(n_-, \alpha, k) \ , \nonumber  \\&& - (\sinh 2\rho_0 + 2 e^{\Phi_0})\Pi(n_+, \alpha, k) \Big] \ .
\end{eqnarray}
where $F$ and $\Pi$'s are the incomplete elliptic integrals of first and third kind respectively with $n_+=\frac{u_1 - u_2}{u_1 + 1}$, $n_-=\frac{u_1 - u_2}{u_1 - 1}$, $\sin\alpha = \sqrt{\frac{u_1 - u}{u_1 - u_2}}$ and $k = \sqrt{\frac{u_1 - u_2}{u_1 - u_3}}$.
Also the spin $S$ and the energy-spin $E-\omega S$ relation reduces to,
\begin{eqnarray}
   S &=& \frac{\sqrt{2}NT_1e^{-\Phi_0}\cosh\rho_1 \sqrt{1 - e^{2\Phi_0}}}{\sqrt{(u_1 - u_3)}} \Big[(u_1 - u_3)E(k) - (1 + u_2) K(k) \nonumber \\ && +\frac{(1+u_2)(1+u_3)}{1+u_1} \Pi(n_+,k) \Big]  \label{4.1} \\
	E - \omega S &=& \frac{\sqrt{2}NT_1e^{-\Phi_0} \sqrt{u_1 - u_3}}{ \sinh\rho_1 \sqrt{1 - e^{2\Phi_0}}} \Big[K(k) - E(k)\Big] \nonumber \\ &+& \frac{\sqrt{2}NT_1}{\sqrt{(u_1 - u_3)}\sqrt{1 - e^{2\Phi_0}} \sinh\rho_1} \Big[ \{ \sinh 2\rho_0 + e^{\Phi_0}(2 + u_1 - u_3)\} K(k) \nonumber \\ && - e^{\Phi_0}(u_1 - u_3) E(k) - (\sinh 2\rho_0 + 2e^{\Phi_0}) \Pi(n_+,k) \Big] \ . \label{4.2}
\end{eqnarray}
One can verify this results (\ref{4.1}) and (\ref{4.2}) also follows from (\ref{3.3}) and (\ref{3.4}) under $q=0$ (pure RR flux) condition. 

Once again we take the long string limit, $\rho_1 >> \rho_0$. Under this limit $k \to \frac{\sqrt{1 - e^{2\Phi_0}}}{\sqrt{\omega^2 - e^{2\Phi_0}}}$ and $\frac{u_1 - u_2}{1 + u_1} \to \frac{1}{\omega^2}$, $S$ and $E-\omega S$ becomes,
\begin{eqnarray}
	S &=& \frac{2NT_1 e^{-\Phi_0}\omega (1 - e^{2\Phi_0})}{\sqrt{\omega^2 - e^{2\Phi_0}}} \Big[ \frac{\omega^2 - e^{2\Phi_0}}{(\omega^2 - 1) (1 - e^{2\Phi_0})} E(k)  - K(k) - \frac{(\omega^2 - 1)e^{2\Phi_0}}{\omega^2 (1 - e^{2\Phi_0})} \Pi \Big(\frac{1}{\omega^2},k\Big)\Big] \ , \nonumber \\ E - \omega S &=&  -\frac{2NT_1 e^{\Phi_0}}{\sqrt{\omega^2 - e^{2\Phi_0}}} \Big[(\omega^2 - 1) \Pi \Big(\frac{1}{\omega^2},k\Big) + \frac{\omega^2 - e^{2\Phi_0}}{1 - e^{2\Phi_0}}E(k) - \frac{2\omega^2 - 1 - \omega^2 e^{2\Phi_0}}{1 - e^{2\Phi_0}} K(k)\Big] \ , \nonumber \\ &+& \frac{2NT_1 e^{-\Phi_0} \sqrt{\omega^2 - e^{2\Phi_0}}}{(1 - e^{2\Phi_0})} \Big[K(k) - E(k)\Big] \ .
\end{eqnarray}
Finally under infinite string limit $\omega \to 1$ the above relations reduces to,
\begin{eqnarray}
	S &=& 2NT_1e^{-\Phi_0} \sqrt{1 - e^{2\Phi_0}} \Big[\frac{1}{2(\omega - 1)}E(k) - K(k)\Big] \ , \nonumber \\ E - S &=& \frac{2NT_1 e^{-\Phi_0} (1  + e^{2\Phi_0})}{\sqrt{1 - e^{2\Phi_0}}} \Big[K(k) - E(k)\Big] \ .
\end{eqnarray}
Now using the expansion of $E(k) \approx 1$ and $K(k) \approx \frac{1}{2}[\log 8 + \log(1 - e^{2\Phi_0}) - \log(\omega -1)]$ in the above expression we get, $\log S = -\log(\omega -1)$ and
finally on substituting the D1-string tension, we get,
\begin{equation}
E - S = \frac{N\sqrt{\lambda}e^{-\Phi_0}[1 + e^{2\Phi_0}]}{\pi g_s\sqrt{1 - e^{2\Phi_0}}} \left[\frac{1}{2}\log S + \cdots\right] \label{4.3}
\end{equation}
One can also verify that this relation (\ref{4.3}) directly follows form (\ref{3.5}). But, the surprising fact about the relation (\ref{4.3}) is that the scaling constant is exactly same as (\ref{3.1}). Thus under infinite string limit the mixed flux energy-spin relation is somehow behaving like the pure RR flux energy-spin relation along with the log divergence in the energy-spin relation. 

	\subsection{Case of $q=1$: Pure NS-NS flux} 
	Here we will consider the background supported by pure NS-NS flux and absence of any R-R flux. This we can study by taking $q=1$. 
	The string action for pure NS-NS flux is 
	\begin{equation}
	S=-T_1 e^{-\Phi_0} \int d\xi^0 d\xi^1 \sqrt{\rho^{\prime 2}~(\dot t^ 2\cosh^2\rho - \dot{\theta}^2\sinh^2\rho)+\theta^{\prime 2}~ \dot t^ 2 ~\sinh^2\rho}
	\end{equation}
	The equation of motion for $\theta$ gives $\partial_1\rho$ to be
	
	\begin{equation}
	\partial_1\rho = \frac{\sinh\rho  \sqrt{4\sinh^2\rho  - \sinh^2 2\rho_0}}{\sinh 2\rho_0 \sqrt{\cosh^2\rho - \omega^2\sinh^2\rho}} \ ,
	\end{equation}
	where ~~$\sinh 2\rho_0 = \frac{2C}{T_1 e^{-\Phi_0}}$. It appears from the above relation that the position of the cusps remains unchanged, as $\partial_1\rho$ diverges at this point and the value of the minimun is simplified considerably to $\rho_{min} = \sinh^{-1}[\frac{\sinh 2\rho_0}{2}]$ which we get by equating $\partial_1\rho=0$. So,the value of $\rho $ varies from a minimum value $\rho=\rho_{min}$ to maximum value $\rho=\rho_1$. 
	Now solving the worldsheet coordinates we get, 
	  \begin{eqnarray}
	 \xi^1 &=& \sinh 2\rho_0 \int \frac{\sqrt{\cosh^2\rho - \omega^2\sinh^2\rho}}{\sinh\rho \sqrt{4\sinh^2\rho - \sinh^2 2\rho_0}}d\rho \nonumber \\ &=& \frac{\sinh 2\rho_0}{\sqrt{2}\sqrt{u_1 + 1} \sinh\rho_1} ~~\Big[ \Pi ~\Big(\frac{u_1 - u_{min}}{u_1 - 1},\alpha, k\Big) - F(\alpha,k)~~\Big] \ .
	 \end{eqnarray}\\
	 where K(k) and $\Pi(n,k)$ are the complete elliptic integrals of the first and third kind respectively, with $ \sin\alpha=\sqrt{\frac{u_1-u}{u_1-u_{min}}},n=\frac{u_1 - u_{min}}{u_1 - 1}$, where $u_{min} = \frac{1+u_0^2}{2}$ and $k = \sqrt{\frac{u_1 - u_{min}}{u_1 + 1}}$. 

	
	
	
	
	In this case the conserved charges energy $E$, spin $S$ and the energy-spin relation $E-\omega S$ becomes,	
	\begin{eqnarray}
	E &=& \frac{\sqrt{2}NT_1e^{-\Phi_0}\sinh\rho_1}{\sqrt{(u_1 + 1)}} \Big[ (u_1 + 1) E(k) - \frac{\omega^2 \sinh^2 2\rho_0}{2} K(k) \Big] \ , \nonumber \\
	S &=& \frac{\sqrt{2}NT_1 e^{-\Phi_0} \cosh\rho_1}{\sqrt{(u_1 + 1)}} \Big[ (u_1 + 1)E(k) - (u_{min} + 1)K(k) \Big] \ , \nonumber \\ E - \omega S &=& \frac{\sqrt{2}NT_1 e^{-\Phi_0}\sqrt{u_1 + 1}}{\sinh\rho_1} \Big[K(k) - E(k)\Big] \ .
	\end{eqnarray}
		
	
	
	 
	 
	Now one can check under infinite string limit, that is for $\rho_1 >> \rho_0$ and $\omega \to 1$, the energy-spin relation becomes, 
	

 
\begin{eqnarray}
     E-S = \frac{N\sqrt{\lambda} e^{-\Phi_0}}{\pi g_s}\left[\frac{1}{2}\log S+\cdots\right] \label{5.1}
\end{eqnarray}
 Once more one can verify that (\ref{5.1}) will directly follows from (\ref{3.5}) under the limit $q \to 1$. Note that the above dispersion relation also consists of the well known logarithmic divergence in its leading order which is similar to the GKP string when we put $N=2$. However, the constant scaling factor is similar to (\ref{1.2}) except for a dilaton factor.

\begin{figure}[t]
{\includegraphics[scale=0.4]{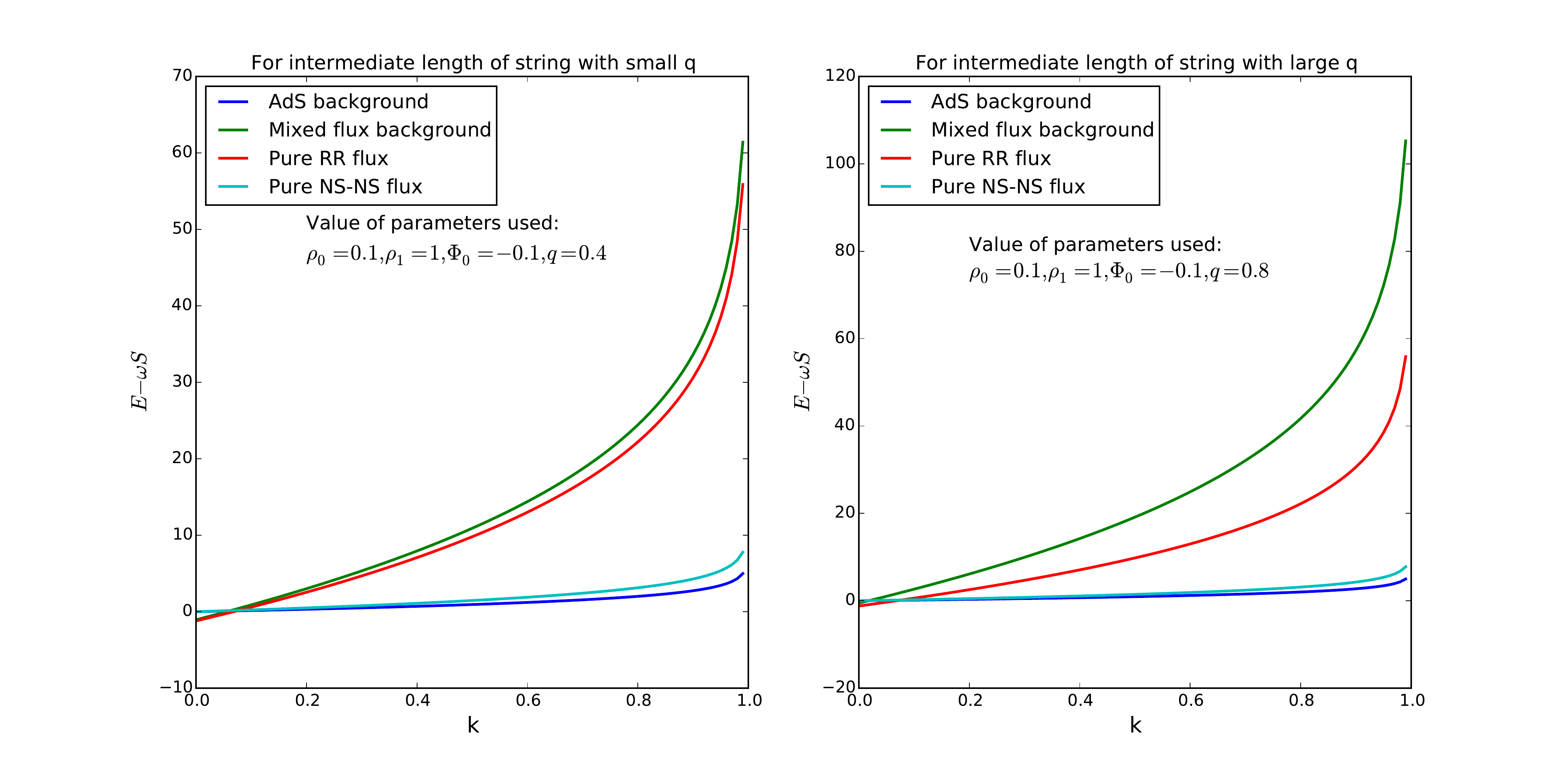}}
\caption{In this figure the spin-energy relation $E-\omega S$ has been plotted against $k$ for different backgrounds under intermediate length of the string. It seems that the pure NS-NS flux spin-energy relation following the spin-energy relation of $AdS$ and these curves are independent of the NS-NS flux. For low value of NS-NS flux ($q=0.4$) pure RR flux spin-energy relation is following the spin-energy relation of mixed-flux background, but for high value of NS-NS flux ($q=0.8$) they follow quite different tracks.}\label{fig5}
\end{figure}

\begin{figure}[t]
{\includegraphics[scale=0.4]{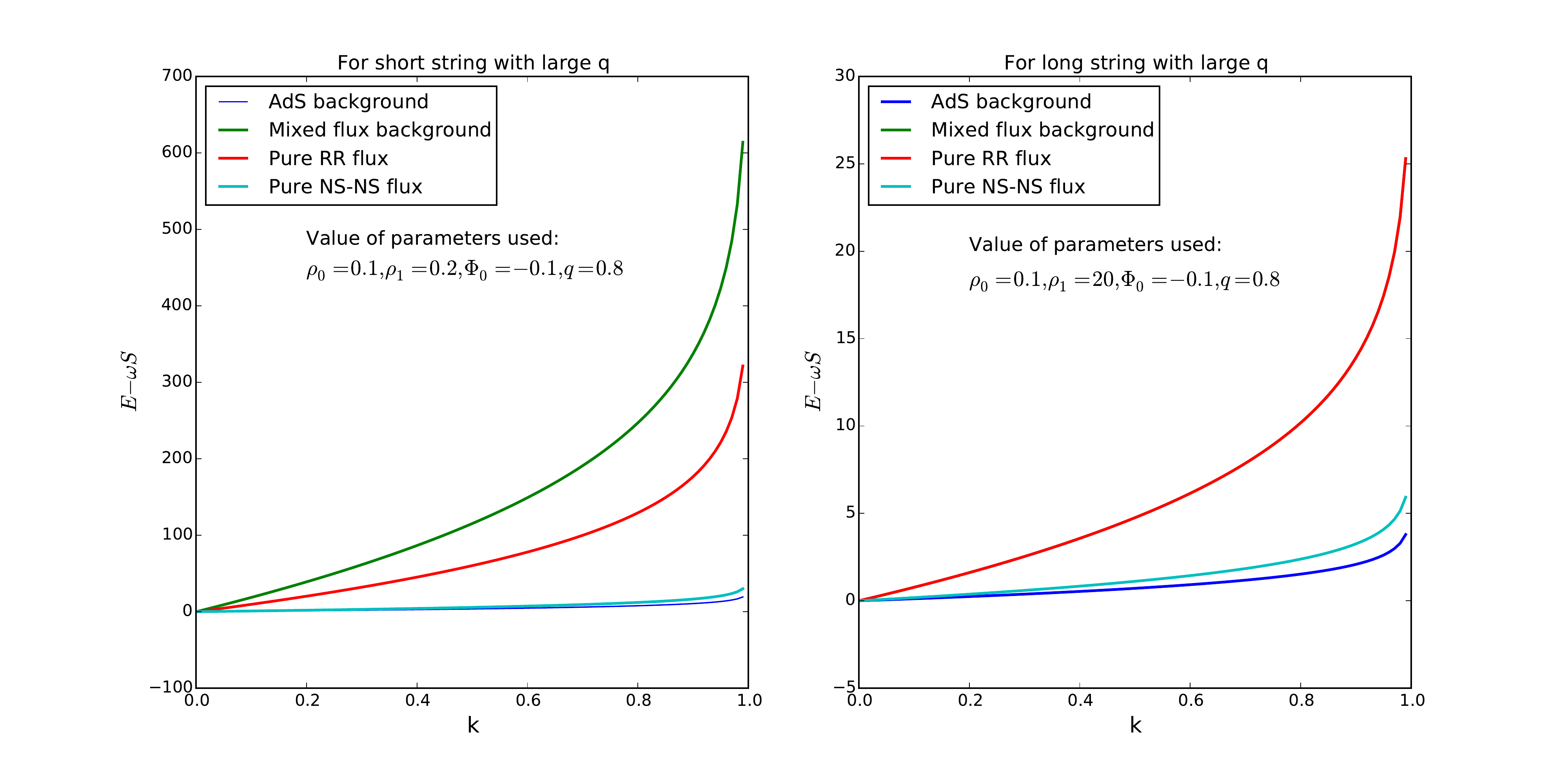}}
\caption{For short string limit the spin-energy relation of pure RR flux background follows quite different curve than the mixed-flux background for high value of NS-NS flux ($q=0.8$). But, for long string limit they seems to converge each other. The energy-spin relation for pure NS-NS background and $AdS$ background seems to be independent of fluxes and length of the string.}\label{fig6}
\end{figure}

Now we are in a situation to compare the enegry-spin relations we have obtained for different cases. These comparison has been presented in Figure \ref{fig5} and Figure \ref{fig6} under different string lengths as before. 
(i) For intermediate string length as shown in Figure \ref{fig5}, it has been found that the energy-spin relation for pure NS-NS flux and $AdS$ background follow similar kind of curves. While the energy-spin relation of pure RR flux and mixed-flux background follows similar kind of curves for lower value of NS-NS flux. One can also note that the first kind of curves (pure NS-NS and $AdS$) start form zero increases slowly until $k=0.99$, while the second kind of curves (pure RR and mixed flux) start with a negative value crossing the first kind of curves increases much rapidly untill $k=0.99$. As usual it is also clear the curves for pure RR flux, pure NS-NS flux and $AdS$ background is independent of $q$, however the curve for mixed-flux background depends on $q$ and increases much rapidly with higher value of $q$.
(ii) Under short string limit the curves of pure NS-NS flux and $AdS$ background doesn't change much but the curve of RR flux increases rapidly and the curve of mixed flux background increases more rapidly as can be seen in Figure \ref{fig6}. Under long string limit rapidity of increasing of these curves are much less (lesser than the intermediate length of string) and interestingly the mixed flux curve converges to the pure RR-flux case background. One can also note the starting point of all the curves converge for both the limits.
 \section{Conclusions and outlook}
 In this paper, by using the DBI action, we have studied a spinning D1-string in $AdS_3$ background with both NS-NS and R-R fluxes. First, we have studied a folded D1-string configuration in pure $AdS_3$ and have calculated the energy spin dispersion relation of such a string. The energy has been shown to be a function of the spin with the logaritmic behavior similar to that of GKP string. The D-string spiky profiles matches with the spiky strings that appears in AdS/CFT.  

 Next, we have studied the D1-string in $AdS_3$ background with mixed three fluxes and investigated the effect of both NS-NS and RR fluxes in our solution. We have found although the position of spike does not depend on these fluxes, both of them are responsible for shifting the position of the valley. A closer look at the spiky D-string profile reveals the effect of mixed fluxed compared to the pure AdS spikes and the fate of such strings in the presence of NS-NS flux. We have depicted them pictorially. 
 In this connection, we have seen $\partial_1\rho=0$ gives two roots in mixed flux background instead of one as in pure $AdS_3$. Although both of these zeros depend on the fluxes, our intuition is that it is the RR flux which is responsible for these two zeros of $\partial_1\rho$ instead of one. Finally, while looking at D1-string profile and conserved charges we found that these quantities can not be expressed in terms of the known elliptic integrals and we have found that NS-NS flux is responsible for this complexity. 
 However, we could express these quantities in terms of some generalised integrals as defined in Appendix A. We have also checked that these generalised integrals reduce to known elliptic integrals of different kinds in the limit $q=0$. To study the behaviour of these generalised integrals we have plotted them by integrating numerically and investigated the effect of the NS-NS fluxes for different lengths of the string. It has been found that the NS-NS flux have significant contibution for intermediate length of string. Under short string limit contribution of the NS NS flux is not so significant while under long string limit these generalised integrals found to converge to elliptic integrals, this might be due to the fact that the long string limit is shielding the effect of NS-NS flux on the generalised integrals. It will be interesting to study further properties of these generalised integrals and we hope they will appear in some other physical situations.

Analytically the generalised integrals is found to reduce to known elliptic integrals of different kind only under infinite string limit and under this limit, the logarithmic divergence of the leading order in the energy-spin relation has been found which is consistent with GKP string solution. It seems that the individual contributions of both NS-NS and RR fluxes only appear in the scaling coefficient of energy-spin relation (\ref{3.5}) but these contributions actually cancel leading to (\ref{3.1}). We have noted that the scaling coefficient of the $(E-S)$ relation in the long string limit, doesnot depend on the amount of flux $q$ in the leading order. This result appears to be slightly awkward as similar N spike solutions obtained from the F-string in the presence of mixed flux was dependent on $q$ \cite{Banerjee:2019puc}. As F-string and D-string are related by S-duality symmetry, one would have expected a similar result in case of D-string. However, we note that legitimately, there is no obvious relation between these two solutions since S- duality symmetry relation does not apply to classical string solutions. It will certainly be interesting to look for the next to leading order term in the dispersion relation to check whether the effect of the flux is evident.

We have further studied the fate of such solutions in the presence of pure NS-NS and pure RR fluxes. Again we have found in the leading order, energy-spin relation have the log divergence in long string limit for both the cases. A comparison of the energy-spin relation for different background has been shown in Figure \ref{fig5} and \ref{fig6}. It has been found that under long string limit, the mixed flux energy-spin relation converges to pure RR flux energy spin relation. We guess this might be due to the fact that under long string limit the effect of the NS-NS flux is negligible (as $q \to 0$) and the mixed flux energy-spin relation is behaving like pure RR flux energy-spin relation. However, the energy-spin relation of pure RR flux (\ref{4.3}) and mixed flux case (\ref{3.5}) becomes equal only under infinite string limit analytically.  Also one can note that the scaling constant of (\ref{5.1}) is same as the scaling constant of (\ref{1.2}) except for a constant dilaton factor. 

One can generalize the results of the present paper to more generic rotating and pulsating D-string in this mixed flux background. The generic extended string might see the effect of flux at the leading order and it might be useful in the study of the dual theory if exists. Further, it will be interesting to study the $(p,q)$ string dynamics in this mixed flux background and see the explicit dependence of the fluxes in the dispersion relation among charges. We will get into some studies along these lines in future.

\section*{ Acknowledgements} We would like to thank Arkady A. Tseytlin and Aritra Banerjee for some useful discussions.   

\section*{Appendix A}
	We define some incomplete generalised integrals as follows:
\begin{equation}
    F_1(\varphi,q,k) = \int_0^{\varphi} \frac{\sqrt{1-n_+\sin^2\theta} d\theta}{\sqrt{1-p^2\sin^2\theta} \sqrt{1-k^2\sin^2\theta}} \ ,
\end{equation}
\begin{equation}
    F_2(\varphi,q,k) = \int_0^{\varphi} \frac{\sqrt{1-p^2\sin^2\theta} d\theta}{\sqrt{1-n_+\sin^2\theta} \sqrt{1-k^2\sin^2\theta}} \ ,
\end{equation}
\begin{equation}
    E_1(\varphi,q,k) = \int_0^{\varphi} \frac{\sqrt{1-n_+\sin^2\theta} \sqrt{1-k^2\sin^2\theta} d\theta}{\sqrt{1-p^2\sin^2\theta}} \ ,
\end{equation}
\begin{equation}
    E_2(\varphi,q,k) = \int_0^{\varphi} \frac{\sqrt{1-p^2\sin^2\theta} \sqrt{1-k^2\sin^2\theta} d\theta}{\sqrt{1-n_+\sin^2\theta}} \ ,
\end{equation}
\begin{equation}
    \Pi(n_+,\varphi,q,k) = \int_0^{\varphi} \frac{ d\theta}{ \sqrt{1-n_+\sin^2\theta} \sqrt{1-p^2\sin^2\theta} \sqrt{1-k^2\sin^2\theta}} \ ,
\end{equation}
\begin{equation}
    I_1(\varphi,q,k) = \int_0^{\varphi} \frac{\sqrt{1-k^2\sin^2\theta} d\theta}{\sqrt{1-n_+\sin^2\theta} \sqrt{1-p^2\sin^2\theta}} \ ,
\end{equation}
and
\begin{equation}
    I_2(\varphi,q,k) = \int_0^{\varphi} \frac{ d\theta}{[1-n_-\sin^2\theta] \sqrt{1-n_+\sin^2\theta} \sqrt{1-p^2\sin^2\theta} \sqrt{1-k^2\sin^2\theta}} \ .
\end{equation}

One can note that for $n_+=p^2$ first two integrals reduce to incomplete elliptic integrals of first kind $F(\varphi,k)$, next two integrals will reduce to incomplete elliptic integrals of second kind $E(\varphi,k)$ and the fifth integral will reduce to the incomplete elliptic integrals of third kind $\Pi(n_+,\varphi,k)$. So, these five integrals can be considered as the generalisation of elliptic integrals of different kinds. The last and second last integrals can be expressed as a combination of incomplete ellipic integrals for $n_+=p^2$.\\

The complete generalised integrals of the above set can be obtained by putting $\varphi = \pi/2$ and are given as follows:
\begin{equation}
    K_1(q,k) = \int_0^{\frac{\pi}{2}} \frac{\sqrt{1-n_+\sin^2\theta} d\theta}{\sqrt{1-p^2\sin^2\theta} \sqrt{1-k^2\sin^2\theta}} \ ,
\end{equation}
\begin{equation}
    K_2(q,k) = \int_0^{\frac{\pi}{2}} \frac{\sqrt{1-p^2\sin^2\theta} d\theta}{\sqrt{1-n_+\sin^2\theta} \sqrt{1-k^2\sin^2\theta}} \ ,
\end{equation}
\begin{equation}
    E_1(q,k) = \int_0^{\frac{\pi}{2}} \frac{\sqrt{1-n_+\sin^2\theta} \sqrt{1-k^2\sin^2\theta} d\theta}{\sqrt{1-p^2\sin^2\theta}} \ ,
\end{equation}
\begin{equation}
    E_2(q,k) = \int_0^{\frac{\pi}{2}} \frac{\sqrt{1-p^2\sin^2\theta} \sqrt{1-k^2\sin^2\theta} d\theta}{\sqrt{1-n_+\sin^2\theta}} \ ,
\end{equation}
\begin{equation}
    \Pi(n_+,q,k) = \int_0^{\frac{\pi}{2}} \frac{ d\theta}{ \sqrt{1-n_+\sin^2\theta} \sqrt{1-p^2\sin^2\theta} \sqrt{1-k^2\sin^2\theta}} \ ,
\end{equation}
\begin{equation}
    I_1(q,k) = \int_0^{\frac{\pi}{2}} \frac{\sqrt{1-k^2\sin^2\theta} d\theta}{\sqrt{1-n_+\sin^2\theta} \sqrt{1-p^2\sin^2\theta}} \ ,
\end{equation}
and
\begin{equation}
    I_2(q,k) = \int_0^{\frac{\pi}{2}} \frac{ d\theta}{[1-n_-\sin^2\theta] \sqrt{1-n_+\sin^2\theta} \sqrt{1-p^2\sin^2\theta} \sqrt{1-k^2\sin^2\theta}} \ .
\end{equation}

\end{document}